\documentclass[letterpaper]{article} 
\usepackage{aaai24}  
\usepackage{times}  
\usepackage{helvet}  
\usepackage{courier}  
\usepackage[hyphens]{url}  
\usepackage{graphicx} 
\usepackage{natbib}  
\usepackage{caption} 
\usepackage{algorithm}
\usepackage{algorithmic}

\usepackage{amsmath,amssymb,amsthm}
\usepackage[mathscr]{euscript}
\newtheorem{problem}{Problem}
\newtheorem{definition}{Definition}

\newcommand{\cA}{\mathscr{A}}
\newcommand{\cC}{\mathscr{C}}
\newcommand{\cM}{\mathscr{M}}
\newcommand{\cS}{\mathscr{S}}
\newcommand{\cT}{\mathscr{T}}
\newcommand{\bbR}{\mathbb{R}}
\newcommand{\bC}{\boldsymbol{C}}
\newcommand{\bM}{\boldsymbol{M}}
\newcommand{\bP}{\boldsymbol{P}}
\newcommand{\bS}{\boldsymbol{S}}
\newcommand{\bb}{{\boldsymbol{b}}}

\newcommand{\bsf}{{\boldsymbol{f}}}

\newcommand{\bh}{{\boldsymbol{h}}}
\newcommand{\bl}{{\boldsymbol{l}}}

\newcommand{\bx}{{\boldsymbol{x}}}
\newcommand{\by}{{\boldsymbol{y}}}
\newcommand{\bz}{{\boldsymbol{z}}}
\newcommand{\rD}{{\mathrm{D}}}

\usepackage[toc, acronym, nopostdot, nonumberlist]{glossaries}
\newacronym{it}{IT}{Information Technology}
\newacronym{apt}{APTs}{advanced persistent threats}
\newacronym{milp}{MILP}{mixed integer linear program}
\newacronym{scada}{SCADA}{supervisory control and data acquisition}
\newacronym{cpes}{CPES}{cyber-physical energy system}
\newacronym{hag}{HAG}{Hybrid Attack Graph}
\newacronym{cwe}{CWE}{Common Weakness Enumeration}
\newacronym{cve}{CVE}{Common Vulnerability Enumeration}
\newacronym{capec}{CAPEC}{Common Attack Pattern Enumeration and Classification}
\newacronym{attack}{MITRE ATT\&CK}{MITRE Adversarial Tactics, Techniques, and Common Knowledge}

\DeclareMathOperator{\diag}{diag}
\DeclareMathOperator{\vul}{Vul}
\DeclareMathOperator{\sys}{S}
\DeclareMathOperator*{\argmin}{arg\,min}

\usepackage{xcolor}

\newcommand{\update}[1]{#1}

\usepackage{newfloat}
\usepackage{listings}

\DeclareCaptionStyle{ruled}{labelfont=normalfont,labelsep=colon,strut=off} 
\floatstyle{ruled}
\newfloat{listing}{tb}{lst}{}
\floatname{listing}{Listing}
\title{Fortify Your Defenses: Strategic Budget Allocation to Enhance Power Grid Cybersecurity\thanks{The research described in this paper is part of the Resilience Through Data Driven, Intelligently Designed Control (RD2C) Initiative at Pacific Northwest National Laboratory (PNNL). It was conducted under the Laboratory Directed Research and Development Program at PNNL, a multiprogram national laboratory operated by Battelle for the U.S. Department of Energy.}}

\author {
    Rounak Meyur,
    Sumit Purohit,
    Braden K. Webb
}
\affiliations {
    Pacific Northwest National Laboratory, Richland\\
    \{rounak.meyur, sumit.purohit, braden.webb\}@pnnl.gov
}

\begin{document}

\maketitle

\begin{abstract}
The abundance of cyber-physical components in modern day power grid with their diverse hardware and software vulnerabilities has made it difficult to protect them from \gls{apt}.
An attack graph depicting the propagation of potential cyber-attack sequences from the initial access point to the end objective is vital to identify critical weaknesses of any cyber-physical system. 
A cyber security personnel can accordingly plan preventive mitigation measures for the identified weaknesses addressing the cyber-attack sequences. 
However, limitations on available cybersecurity budget restrict the choice of mitigation measures. 
\update{We address this aspect through our framework, which solves the following problem: given potential cyber-attack sequences for a cyber-physical component in the power grid, find the optimal manner to allocate an available budget to implement necessary preventive mitigation measures.} 
We formulate the problem as a \gls{milp} to identify the optimal budget partition and set of mitigation measures which minimize the vulnerability of cyber-physical components to potential attack sequences.
We assume that the allocation of budget affects the efficacy of the mitigation measures. 
We show how altering the budget allocation for tasks such as asset management, cybersecurity infrastructure improvement, incident response planning and employee training affects the choice of the optimal set of preventive mitigation measures and modifies the associated cybersecurity risk. 
\update{The proposed framework can be used by cyber policymakers and system owners to allocate optimal budgets for various tasks required to improve the overall security of a cyber-physical system.}
\end{abstract}

\section{Introduction} \label{sec:intro}
An increased reliance on \gls{it} in various aspects of modern life has created a vast ecosystem of interconnected systems, networks and devices~\cite{Bansal2020}.
This provides a large attack surface for cyber adversaries to target, allowing them to gain unauthorized access, steal data, and disrupt operations.
The availability of off-the-shelf hacking tools and malware in the underground market makes their task even easier to an extent, which allows them to initiate complex attacks without the requirement of sophisticated programming expertise~\cite{Liggett2019}.
Sophisticated malware, \gls{apt} and multi-stage cyber attacks involving multiple attack vectors are principal factors, which increase the complexity of these attacks and make them harder to detect and mitigate~\cite{Li2021}.
\update{This requires system owners and cybersecurity personnel to be well aware about the latest vulnerabilities and plan to mitigate them effectively.}

The modern day energy infrastructure is equipped with smart devices, which aid in its monitoring and control.
These devices are an integral part of the \gls{cpes}.
They form the link between the physical power grid and the communication network, allowing system operators to take online decisions and alter system conditions remotely.
However, this comes at a cost of increased vulnerability to cyber attacks where adversaries can gain access to these devices and adversely impact the power grid infrastructure, leading to severe events such as widespread blackout.
A typical \gls{cpes} consists of multiple smart devices interlinked through a communication network.
These smart devices (such as a smart inverter or a protective relay) can be accessed directly by a cyber adversary or via the communication network after a successful intrusion into a centrally situated device (such as a substation automation controller).
Therefore, the goal of cybersecurity personnel is to protect these smart devices (or components) from adversarial cyber intrusions.
From hereon, we use the terms `component' and `smart device' interchangeably.
In this work, we aim to identify an optimal set of preventive cybersecurity measures for each component in the \gls{cpes} in order to reduce the risk of adversarial cyber attacks. 

A bottom-up approach involves evaluating the risk associated with the failure of a component by assessing the loss of power grid resilience or stability, and thereafter, allocating budget towards securing the `critical' components~\cite{ioannis2021}.
On the contrary, a top-down approach focuses on cyber vulnerabilities for a component, possible adversarial techniques used to exploit them, and preventive strategies to avoid them~\cite{sumit2022,dutta2022cyber,sumit2022impact}.
\update{This involves developing and implementing patches for individual vulnerabilities in a timely manner, which has been reported to be almost impossible~\cite{patch}.
One of the major roadblocks responsible for this is the lack of resources required to cover the sheer surface area of diverse hardware and software vulnerabilities.
To this end, an organizational effort is required which addresses the prioritizing the vulnerabilities and allocating available budget based on their priorities.}


The \gls{attack} framework, developed by the MITRE Corporation, serves as a comprehensive and structured database to understand and organize information about cyber threats~\cite{mitre}.
The framework consists of \emph{tactics}, or high-level objectives of an adversary during an attack, and \emph{techniques}, or specific methods and procedures to accomplish their objectives within each tactic.
It also provides detailed information about associated \emph{mitigation} measures, which refer to strategies that organizations can employ to defend against or reduce the impact of specific techniques.
However, information regarding both the cost of implementing the preventive mitigation measures and their efficacy against adversarial techniques are not included in the framework. 
This makes the task of evaluating the investment required to implement a proposed mitigation plan difficult to compute.
At the same time, it is difficult to quantify how the presence or absence of mitigation measures affects the success rate of an adversarial technique.

In this paper, we approach the evaluation of optimal policies to improve a component's cybersecurity in the \gls{cpes} with an aim to alleviate its risk to adversarial threats.
\update{An important aspect of policy formulation is to prioritize the problems to address and partitioning available budget to implement appropriate solutions.
We treat the cybersecurity budget to be representative of the labor/staff hours and associated resources required to implement the different mitigation measures.
Hence, we identify multiple organizational sectors to segregate the mitigation measures based on the skill or number of staff hours required for implementation~\cite{sensors2021}.
Thereafter, our proposed approach evaluates the high priority mitigation measures required to be implemented and the optimal manner of partitioning the cybersecurity budget to achieve this task.
The underlying assumption of our approach is that allocating budget in a particular sector implies prioritizing mitigation measures within that sector.
This improves the overall efficacy of the mitigation against all adversarial techniques.}
The formal problem statement can be stated as follows:
\emph{
Given potential cyber-attack  sequences for a cyber-physical component in the power grid, find the optimal manner to allocate an available labor budget to implement necessary preventive mitigation measures, which reduces the risk to adversary threats.}

\noindent\textbf{Contributions.}~The major contributions of this paper are listed below: (i) we use a top-down approach to define the vulnerability of a \gls{cpes} component based on the adversarial threats and attack sequences to which it is susceptible, 
(ii) we use the \gls{attack} framework to define the efficacy of mitigation measures against adversarial techniques and thereby, propose analytic expressions to evaluate the success rate of both individual techniques and entire attack sequences, 
(iii) we formulate a \gls{milp} by using these analytic expressions to identify the optimal partitions of a given limited budget to improve mitigation measure efficacy and evaluate the optimal mitigation set required to minimize successful adversarial attack sequences on the cyber component.
The proposed holistic framework can be generalized for any cyber-physical system or any component/system with recorded cyber vulnerabilities.

\section{Related Works}
Risk assessment of \gls{cpes} has been studied extensively using various methodologies, where the impact of cyber attacks on specific nodes in the power network is analyzed to identify the resulting damage.
This has been done either through low fidelity simulation frameworks~\cite{anastasis2016, chen2014, georg2013, dorsch2014, quiroz2011} or through high fidelity real time simulation test beds~\cite{thia2022, sridhar2017, voltron2013, stanovich2013}. These frameworks are useful, since they provide means to identify critical components in terms of their impact on the power grid and enable system planners to focus their cybersecurity countermeasures. 
However, recent intrusion reports show complex cyber attack sequences, which utilize the interconnected nature of communication systems in order to gain system-wide access~\cite{mitre}.
This necessitates a top-down approach, which identifies possible cyber attack sequences and recommend countermeasures to prevent them.

\update{Authors in~\cite{nandi2016} propose an interdiction plan by deploying countermeasures at optimal set of edges in an attack graph to minimize losses due to security breaches.
However, this work assumes a deterministic graph, with a $100\%$ or $0\%$ breach success rate along the edges and also considers fixed budget for countermeasures along each edge in the graph}
The \gls{attack} framework has been used in~\cite{sumit2022,dutta2022cyber} to identify vulnerabilities in several \gls{cpes} components and generate attack sequences.
It provides list of mitigation measures to prevent adversarial techniques. 
However, a holistic framework to identify the optimal set of countermeasures to prevent a set of attack sequences is not available in the present literature. 

The authors in~\cite{Li2019} have performed extensive survey to show how investing in various organizational sectors has affected the overall improvement of cybersecurity awareness in several organizations.
The association of various mitigation measures specified in the \gls{attack} framework to the cybersecurity culture of various organizations has been presented in~\cite{sensors2021}.
This allows a holistic approach to address cybersecurity gaps in infrastructure and policies for an organization.
The allocation of resources to improve cybersecurity in an organization is a pertinent problem due to contrasting interest of different individuals~\cite{Srinidhi2015}.
In this regard, the present literature lacks a framework capable of addressing the aspect of allocating budget to different organizational sectors serving a common goal of reducing vulnerability of adversarial cyber attacks. 
This paper aims to address this particular research gap.

\section{Preliminaries}\label{sec:prelim}

\textbf{\gls{attack} framework.}
The \gls{attack} framework~\cite{mitre} serves as a database of cyber attack scenarios and methods undertaken by adversaries for cyber intrusion.
It contains an extensive list of tactics and techniques common to adversarial cyber attacks.
The `tactics' denote adversarial motivation and `techniques' represent the instrumental means of achieving those tactical objectives. Moreover, each technique is associated with a list of mitigation measures such that a particular cyber defense system with a given set of mitigation measures will be able to prevent only their corresponding techniques.
We denote sets of $N_{\cT}$ techniques and $N_{\cM}$ mitigation measures by $\cT$ and $\cM$ respectively and define a mapping $g:\cT\rightarrow\cM$ to identify the set of mitigation measures $g\left(t\right)\subseteq\cM$, which can prevent an adversary from performing technique $t$. 
The pre-image of a mitigation measure $m$ under $g$ provides the induced mapping $g^{-1}:\cM\rightarrow\cT$ as the set of techniques $g^{-1}\left(m\right)$ which can be prevented when $m$ is available in the cyber defense system.

\begin{figure}[tbhp]
    \centering
    \includegraphics[width=0.47\textwidth]{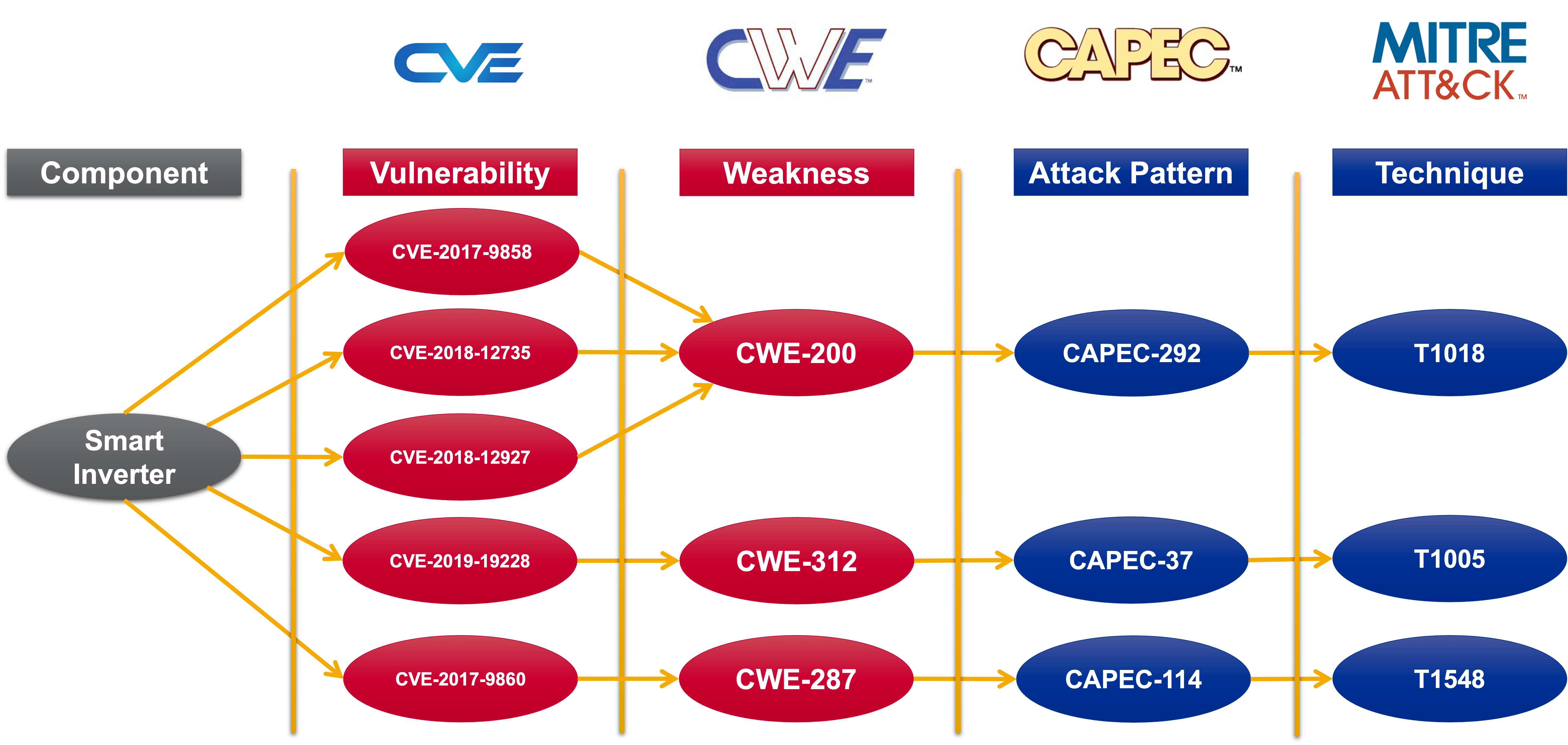}
    \caption{Framework to map between various vulnerability and attack pattern databases. The list of adversarial techniques that might be possible to be executed on a ``smart inverter'' are obtained in this mapping.}
    \label{fig:mapper}
\end{figure}

\noindent\textbf{\gls{hag}.}
The cybersecurity risk assessment of a component in the \gls{cpes} involves identifying possible adversarial techniques that can be performed on it.
To this end, we use a mapping framework, which maps common cyber vulnerabilities in the component to common attack patterns that can be executed on these vulnerabilities. The framework is depicted in 
Fig.~\ref{fig:mapper}, where we provide the name of a component as input and the framework evaluates the vulnerabilities from the \gls{cve} database, identifies respective weaknesses from the \gls{cwe} database, gets attack patterns from the \gls{capec} database and finally maps to the \gls{attack} framework to obtain the list of adversarial techniques. Refer~\cite{dutta2022cyber} for details about each of these database.

However, cybersecurity risk assessment also requires us to identify the sequence of techniques which can be performed on the component.
A \gls{hag} serves as an excellent tool for this purpose.
These are synthetically generated graphs which are created based on past instances of cyber attacks reported in openly available cyber intrusion reports \cite{donald2023hybrid}.
An attack graph describes attack sequences through a set of possible techniques from the \gls{attack} framework. 
The nodes in the graph represent adversarial techniques and the edges depict consecutive techniques used in an attack sequence. 

\begin{figure}[tbhp]
    \centering
    \includegraphics[width=0.47\textwidth]{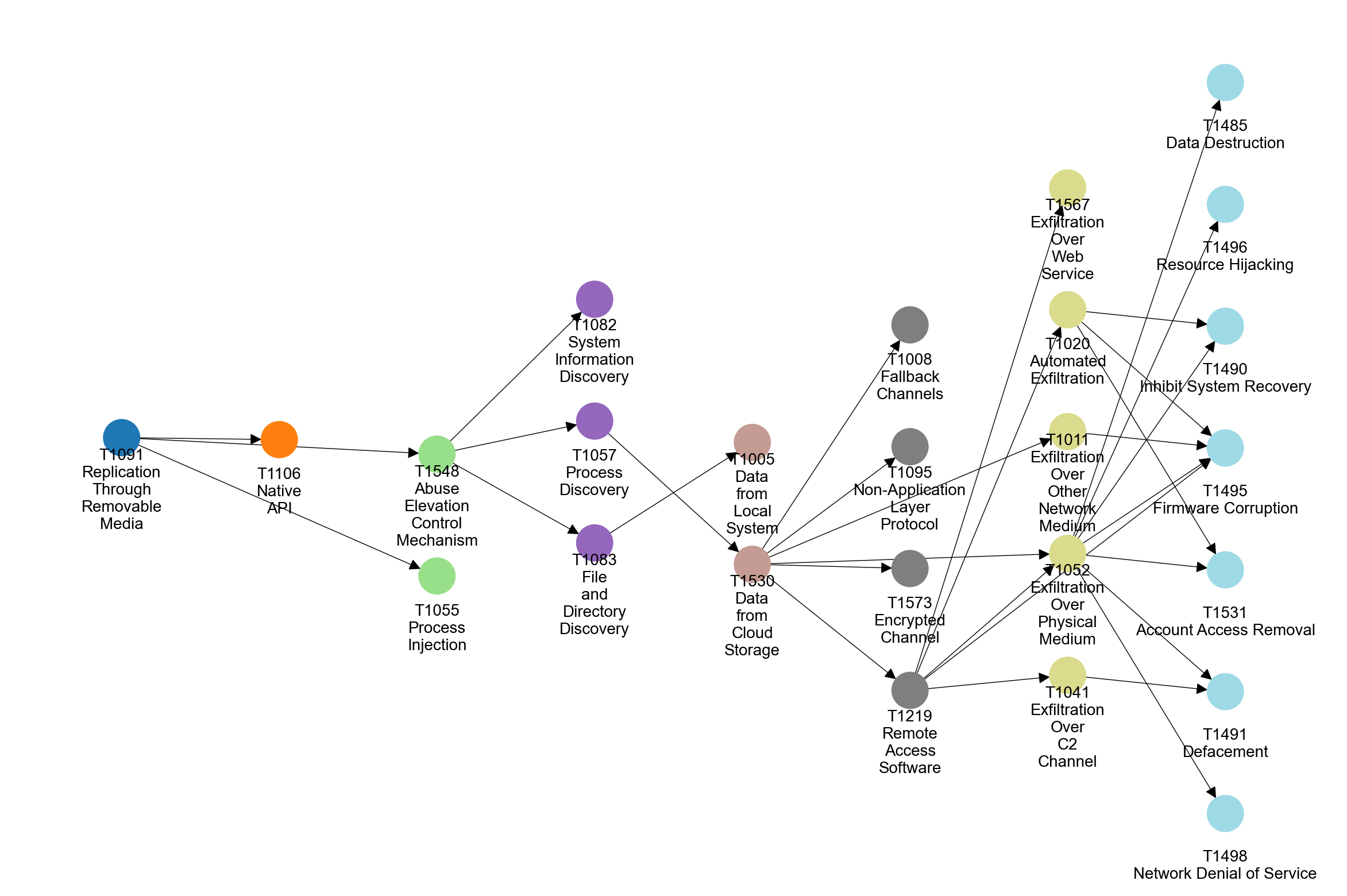}
    \caption{A sample \gls{hag} describing the possible attack sequences that can be performed on a ``smart inverter''. Each node denotes an adversarial technique and nodes with same color denote techniques belonging to the same tactic.}
    \label{fig:hag-inverter}
\end{figure}

A single \emph{attack sequence} is represented as a \emph{path} in the \gls{hag}, describing the progression of techniques. 
We denote an attack sequence of length $n$ as $\cA:=\left\{t_{1}, t_{2}, \cdots, t_n \right\}$. The adversarial techniques $t_{1}, t_{2}, \cdots, t_n$ which comprise the sequence are nodes in the \gls{hag}.
It is important to mention that we assume the following while creating a \gls{hag}: (i) the transition of techniques follow a predefined order of the associated tactics, and (ii) techniques are not repeated.
An example \gls{hag} is shown in Fig.~\ref{fig:hag-inverter}. 

\noindent\textbf{Vulnerability of a component.}
The goal of this work is to harness the \gls{attack} framework to plan mitigation measures for a component.
To this end, we need to define a component's \emph{vulnerability} in the context of the \gls{attack} framework and the \gls{hag} generated for the component.
The generated \gls{hag} provides us with possible sequences of techniques that adversaries could utilize to perform a successful cyber attack.
A planner might seek to choose an optimal set of mitigation measures to minimize the probability of compromising the component through any of the attack sequences identified through the \gls{hag}.
\update{This goal objective requires us to minimize the success probability of each and every attack sequence in the \gls{hag}.}
\update{However, for all practical purposes, reducing these success probabilities to a sufficiently small value is acceptable. We term this objective as minimizing the number of ``highly likely'' attack sequences.}
First, we define what we mean by a ``highly likely'' attack sequence.
Here, we assume that the probability of successful execution of the techniques are independent.
\begin{definition}
A sequence $\cA_l:=\left\{t_{1}, t_{2}, \cdots, \right\}$ is said to be ``highly likely'' if the probability of its successful execution (or success rate $v_l$) exceeds a chosen threshold $\delta > 0$, i.e., 
\begin{equation}
    v_l = \prod_{t\in\cA_l}r_t \geq \delta \quad \iff \quad \sum_{t\in\cA_l}\log{r_t} \geq \log{\delta}
\end{equation}
where $r_t$ is the success rate of technique $t\in\cA_l$.
\label{def:highlike}
\end{definition}
\begin{definition}
The vulnerability of the component $\rD$ with a set of mitigation measures $\cM_{\sys}\subseteq\cM$ is given by the fraction of ``highly likely'' attack sequences
\begin{equation}
    \vul\left(\rD,\cM_{\sys},\delta\right) = \frac{N_{\rD}\left(\cM_{\sys}, \delta \right)}{N_{\rD}}\label{eq:vul}
\end{equation}
where $\delta$ denotes the threshold success rate for a technique to be ``highly likely'' and $N_{\rD}\left(\cM_{\sys},\delta\right)$ denotes the number of ``highly likely'' attack sequences.
\label{def:vul}
\end{definition}

\begin{table}[tbhp]
\centering
\caption{Sets and set elements used}
\label{tab:symbols-sets}
\begin{tabular}{ll}
\hline
\textbf{Symbol} & \textbf{Description}                                \\ \hline
$\cT$           & Set of all cyber adversary techniques               \\
$\cM$           & Set of all mitigation measures                      \\
$\cC$           & Set of cybersecurity budget sectors                 \\
$\cM_{\sys}$    & Set of selected mitigation measures                 \\
$\cS_{\rD}$     & Set of attack sequences for device $\rD$            \\
$N_{\cT}$       & Number of adversary techniques                      \\
$N_{\cM}$       & Number of mitigation measures                       \\
$N_{\rD}$       & Number of attack sequences in $\cS_{\rD}$           \\
$N_{\cC}$       & Number of cybersecurity budget sectors              \\
$t$             & An adversarial technique in set $\cT$               \\
$m$             & An adversarial technique in set $\cM$               \\
$\cA$           & An attack sequence                                  \\ \hline
\end{tabular}
\end{table}

\section{Proposed Approach}\label{sec:problem}
The goal of cybersecurity planning is to identify which mitigation measures to implement to reduce the vulnerability of the cyber-physical component under consideration.
We define the \emph{optimal defender problem} as follows.
\begin{problem}[Optimal Defender Problem]\label{prob:defend}
Given limited budget to enhance cybersecurity of a component $\rD$, find the optimal set of mitigation measures $\cM_{\sys}$ to minimize its vulnerability $\vul\left(\rD,\cM_{\sys}\right)$ for a given set of attack sequences $\cS_{\rD}$.
\end{problem}
The main challenge arises when evaluating the cost of implementing each mitigation measure---either in terms of monetary investment or required time commitment. Furthermore, the efficacy of each mitigation measure against the adversarial techniques is usually unknown.
However, we note that allocating additional budget generally improves the efficacy of mitigation measures.
Therefore, we formulate the problem in a way to to address how to allocate a limited cybersecurity budget to reduce component vulnerability. 
We state the \emph{strategic cybersecurity budget allocation} problem as follows:
\begin{problem}[Cybersecurity Budget Allocation Problem]\label{prob:defend-2}
Given a limited budget to enhance the cybersecurity of a component $\rD$, find the optimal way to partition the budget in order to improve efficacy of a set of mitigation measures $\cM_{\sys}$ and thereby minimize the vulnerability $\vul\left(\rD,\cM_{\sys}\right)$ of the component with a given set of attack sequences $\cS_{\rD}$.
\end{problem}

\noindent\textbf{Budget allocation.}
Following~\cite{sensors2021}, we categorize the mitigation measures into the following overlapping sectors (or categories): asset management, business continuity, access and trust, operations, defense, security governance and employee training.
We partition the entire available budget into the above $N_{\cC}$ sectors. 
Let $b_j$ denote the portion of budget assigned to the $j^{th}$ category and $\sum_j {b_j} = 1$.
Further, we define matrix $\bC\in\left\{0,1\right\}^{N_{\cM}\times N_{\cC}}$ such that $C_{ij}=1$ if the $i^{th}$ mitigation is included in the $j^{th}$ category, otherwise $C_{ij}=0$.

The underlying assumption is that allocating budget improves mitigation measure efficacy, i.e., the probability that a mitigation measure successfully prevents a technique.
In our case, we also assume that for a given mitigation, this probability is uniform for all associated techniques.
A mitigation measure $m_i\in\cM$ belongs to one or more of the $N_{\cC}$ sectors. 
We assume that in order to improve the efficacy of a mitigation, the budget must be allocated to all of the associated sectors to which it belongs.
Therefore, we compute the fractional budget $f_i$ allocated for improving efficacy of mitigation measure $m_i$ as the weighted sum of the category budgets, 
\begin{equation}
    f_i = \frac{\sum_{j=1}^{N_{\cC}}C_{ij}b_j}{\sum_{j=1}^{N_{\cC}}C_{ij}}
    \label{eq:frac-budget}
\end{equation}
Let $\bsf$ be the $N_{\cM}$-length vector obtained by stacking the $f_i$ entries for all mitigation measures.
The matrix version of (\ref{eq:frac-budget}) is written as $\bsf = \diag\left(\bC \boldsymbol{1}\right)^{-1} \bC \bb$, where $\boldsymbol{1}$ is a vector of $1$s.

\begin{figure}[tbhp]
    \centering
    \includegraphics[width=0.46\textwidth]{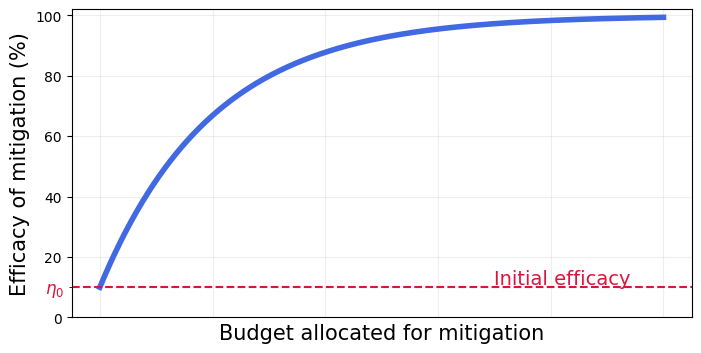}
    \caption{Improvement in mitigation efficacy with increase in allocated budget.}
    \label{fig:efficacy}
\end{figure}

Let $\eta_{i,0}\in\left[0,1\right)$ denote the initial efficacy of the $i^{th}$ mitigation.
\update{This depends on the efficacy of the mitigation measures already in place -- for example, the strength of firewall.}
We define an exponential improvement in the efficacy with increase in the cybersecurity labor budget, such that it is asymptotic to value of $1.0$ based on the following expression:
\begin{equation}
    \eta_i = 1 - \left(1-\eta_{i,0}\right)e^{-\lambda f_i} \label{eq:improve-eta}
\end{equation}
where $\eta_i$ is the improved efficacy 
and $\lambda$ is a suitable scaling factor to relate the improvement in efficacy to the overall budget allocation.
Fig.~\ref{fig:efficacy} shows the exponential improvement in mitigation efficacy for $\lambda=0.1$.
\update{An exponential relation mimics the most natural behavior of diminishing returns on investments and has been used in similar models~\cite{dimret}.}

In practice, the parameter $\lambda$ denotes the organization's efficiency in utilizing the allocated budget to improve the overall cybersecurity.
A high $\lambda$ means a higher rate of improvement in efficacy for a given increase in budget allocation factor.
Further, note that for a given $\lambda$, the maximum improvement in efficacy of mitigation $i$ occurs when budget is allocated to all the associated sectors. 

\noindent\textbf{Success rate of techniques.}
Let $p_{i,k}$ be the probability that technique $t_k\in\cT$ is avoided by a mitigation measure $m_i\in\cM$.
Note that $p_{ik}=0$ for all techniques $t_k\in\cT$ which cannot be mitigated by $m_i$, i.e., $t_k \notin g^{-1}\left(m_i\right)$.
Based on the assumption mentioned in the previous section, we have
\begin{equation}
    p_{i,k} = 
    \begin{cases}
    \eta_i, \quad\quad \textrm{if } t_k \in g^{-1}\left(m_i\right) \\
    0, \quad\quad \textrm{otherwise}
    \end{cases}
    \label{eq:prob-define}
\end{equation}

\begin{figure}[tbhp]
    \centering
    \includegraphics[width=0.47\textwidth]{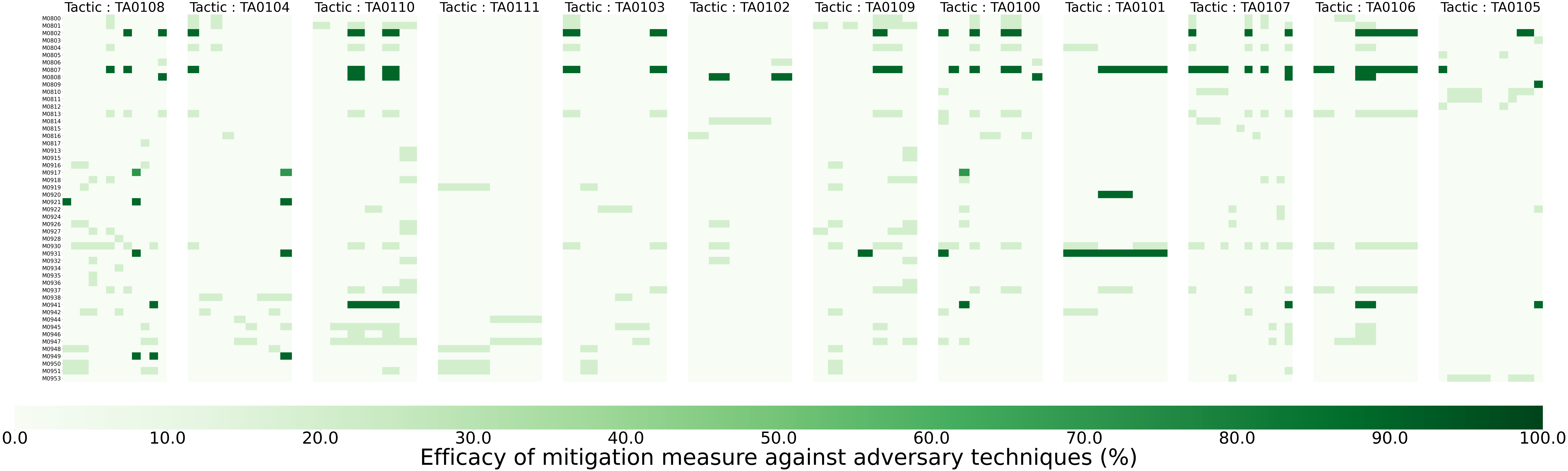}
    \caption{Mitigation-technique relation matrix with efficacy values for each mitigation measure against adversarial techniques.}
    \label{fig:mit-tech-mat}
\end{figure}

Let $\bM\in\left\{0,1\right\}^{N_{\cM}\times N_{\cT}}$ denote the mitigation-technique relation matrix. The entry $M_{ik}$ along the $i^{th}$ row and $k^{th}$ column of $\bM$ is $1$ if the technique $t_k\in\cT$ is mitigated by mitigation measure $m_i\in\cM$; otherwise the entry is $0$. 
This is constructed from the \gls{attack} framework. 
Fig.~\ref{fig:mit-tech-mat} shows the matrix through a heat-map where the rows denote the mitigation measures and columns represent techniques. 
The techniques for each tactic are grouped together. The opacity of every element in the matrix shows the efficacy of the mitigation measure against the adversarial technique.
We call this matrix as the \emph{mitigation profile}.

We want to identify the set of mitigation measures $\cM_{\sys} \subseteq \cM$ which would reduce the vulnerability of the component $\rD$ with possible attack sequences listed in $\cS_{\rD}$.
Let $x_i\in\left\{0,1\right\}$ denote the absence/presence of mitigation measure $m_i$ in the set $\cM_{\sys}$.
Note that the mitigation measures which are not present in the cyber system cannot affect the success rate of an adversarial technique.
Using (\ref{eq:prob-define}) we can write
\begin{equation}
    p_{ik} = \eta_i x_i M_{ik} \label{eq:prob-efficacy}
\end{equation}
Observe that $1-p_{ik}$ indicates the probability that technique $t_k$ is not avoided by mitigation measure $m_i$. 
We consider each event of technique $t_k$ being avoided by mitigation $m_i,~\forall m_i\in\cM$ to be independent.
Therefore, the success rate of a technique $t_k$ in the cyber system with a set of mitigation measures $\cM_{\sys}$ is computed as
\begin{equation}
    r_k = \prod_{i,m_i\in\cM_{\sys}}{\left(1 - p_{ik}\right)} = \prod_{i=1}^{N_{\cM}}{\left(1 - x_i M_{ik} \eta_{i}\right)} \label{eq:success_rate}
\end{equation}
The logarithm of (\ref{eq:success_rate}) is computed as
\begin{equation}
    L_k = \log\left(r_k\right) = \sum_{i=1}^{N_{\cM}}\log\left(1 - x_i M_{ik} \eta_i\right) \label{eq:log_success_rate}
\end{equation}
It is interesting to note that 
\begin{equation}
    \log\left(1-x_i M_{ik} \eta_i\right) = 
    \begin{cases}
        \log\left(1-\eta_{i}\right) \quad~~  \textrm{if}~ x_i=1, M_{ik}=1\\
        0 \quad\quad\quad\quad\quad\quad ~ \textrm{otherwise}
    \end{cases}
\end{equation}
which helps us simplify the log success rate of a technique $t_k$ as
\begin{equation}
    L_k = \log\left(r_k\right) = \sum_{i=1}^{N_{\cM}}x_{i}M_{ik} \log\left(1-\eta_i\right)\label{eq:log-simple}
\end{equation}
Using (\ref{eq:improve-eta}) and (\ref{eq:log-simple}) we have
\begin{equation}
    \log\left(r_k\right) = \sum_{i=1}^{N_{\cM}} x_i M_{ik} \left[\log{\left(1 - \eta_{i,0}\right)} - \lambda f_i \right] \label{eq:log_success_rate_mod}
\end{equation}
Let $\bx$ be the $N_{\cM}$-length vector constructed by stacking the $x_i$ for all mitigation measures.
We define matrix $\bP\in\mathbb{R}^{N_{\cM}\times N_{\cT}}$ with element $P_{ik}=M_{ik} \log\left(1-\eta_{i,0} \right)$ along the $i^{th}$ row and $k^{th}$ column.
Note that $P_{ik}\leq 0$ for all $i,k$ since it is the logarithm of fractional values.
The logarithm of the success rate of technique $t_k$ can therefore be computed as
\begin{equation}
    \log\left(r_k\right) = \left[\bP^T\bx\right]_k - \lambda \left[\bM^T \diag(\bsf) \bx \right]_k
    \label{eq:log_prob}
\end{equation}
where $\left[\bz\right]_k$ denotes the $k^{th}$ element of vector $\bz$.

Next, we evaluate the success rate of an attack sequence. 
Recall that an attack sequence is a list of techniques.
We define the attack sequence and technique relation matrix $\bS\in\left\{0,1\right\}^{N_{\rD}\times N_{\cT}}$ where the entry $S_{lk}$ along the $l^{th}$ row and $k^{th}$ column is $1$ if the technique $t_k$ is present in the attack sequence $\cA_l\in\cS_{\rD}$ and $0$ otherwise.
The logarithm of success rate of an attack sequence can be computed as
\begin{equation}
\log\left(v_{l}\right) = \sum_{t_k\in\cA_l}{\log{r_k}} = \sum_{k=1}^{N_{\cT}}{S_{lk}\log{r_k}} 
\end{equation}
We can stack $\log\left(v_{l}\right)$ for all the attack sequences to an $N_{\rD}$-length vector $\bl$, which can be expressed as
\begin{equation}
    \bl = \bS  \bP^T \bx - \lambda \bS \bM^T \diag(\bsf) \bx 
    \label{eq:success-seq}
\end{equation}
We can formulate Problem~\ref{prob:defend-2} as follows
\begin{subequations}
\begin{align}
    \min_{\bb, \bx}\quad & \vul\left(\rD,\cM_{\sys}\right) \\
    \textrm{s.to}\quad & \bl = \bS  \bP^T \bx - \lambda \bS \bM^T \diag(\bsf) \bx \label{seq:bilinear}\\
    & \bsf = \diag\left(\bC \boldsymbol{1}\right)^{-1} \bC \bb\\
    & \mathbf{1}^T\bb = 1,~~\bb \geq 0
\end{align}
\label{eq:optim-main}
\end{subequations}

\begin{table}[tbhp]
\centering
\caption{Matrices, vectors and variables used}
\label{tab:symbols-vars}
\begin{tabular}{ll}
\hline
\textbf{Symbol} & \textbf{Description}                                        \\ \hline
$\bC$           & Mitigation \& budget category relation matrix              \\
$\bM$           & Mitigation \& technique relation matrix          \\
$\bS$           & Attack sequence \& technique relation matrix     \\
$\bx$           & Vector of mitigation measure indicator   \\
$\by$           & Vector of attack sequence indicator \\
$\bsf$          & Vector of mitigation specific budget partitions             \\
$\bb$           & Vector of cybersecurity budget partitions                   \\
$\bl$           & Vector of log of attack sequence success rate         \\
$\eta_i$        & Efficacy of mitigation $m_i$                                \\
$r_k$           & Success rate of technique $t_k$                             \\
$v_l$           & Success rate of attack sequence $\cA_{l}$\\
$\lambda$       & Skill level of defender                                     \\ \hline
\end{tabular}
\end{table}

\subsection{Proposed Optimization Framework}\label{ssec:opt-defend}
First, we define variable $h_i=f_ix_i$ to get rid of the bi-linear product term in the expression of (\ref{seq:bilinear}). The corresponding $N_{\cM}$-length vector obtained by stacking them is denoted by $\bh$. Since $x_i\in\{0,1\}$ and $0 \leq f_i\leq 1$, we can write the following inequalities
\begin{subequations}
\begin{align}
    & \bh \leq \bx \\
    & \bh \geq \boldsymbol{0} \\
    & \bh \leq \bsf \\
    & \bh \geq \bx - \left( \boldsymbol{1} - \bsf \right)
\end{align}
\label{eq:milp-1}
\end{subequations}
Note that when $x_i=0$, we obtain the equality $h_i=0$ from the first two inequalities, and when $x_i=1$, we obtain $h_i=f_i$ from the last two inequalities.

We use the definition of vulnerability described through (\ref{eq:vul}), which leads us to a \gls{milp} as discussed below.
We define the binary variable $y_l\in\left\{0,1\right\}$ to denote whether sequence $\cA_l\in\cS_{\rD}$ is ``highly likely'' or not. 
Mathematically,
\begin{equation}
    y_l = 
    \begin{cases}
    1, \quad\quad \mathrm{if}~\left[\bS \left( \bP^T \bx - \lambda \bM^T \bh \right) \right]_l \geq \log{\delta}\\
    0, \quad\quad \mathrm{otherwise}
    \end{cases}
    \label{eq:seq-cases}
\end{equation}
Define $\delta' = \log{\delta}$. 
We can rewrite (\ref{eq:seq-cases}) for all sequences $\cA_l\in\cS_{\rD}$ with a large positive constant $K$ using the following inequalities
\begin{subequations}
\begin{align}
    & K \by \geq \bS\bP^T\bx - \lambda \bS \bM^T \bh - \delta'\mathbf{1} \\
    & K \left(\mathbf{1} - \by \right) \geq  \delta'\mathbf{1} - \bS\bP^T\bx + \lambda \bS \bM^T \bh
\end{align}
\label{eq:milp-2}
\end{subequations}
Our aim is to minimize the number of ``highly likely'' attack sequences. 
We can therefore write the optimization problem as
\begin{subequations}
\begin{align}
    \bx^{\star} = \argmin_{\bb\in\bbR^{N_{\cC}}, \bx \in \left\{0,1\right\}^{N_{\cM}}} & \mathbf{1}^T\by \\
    \textrm{s.to}\quad & \bsf = \diag\left(\bC \boldsymbol{1}\right)^{-1} \bC \bb\\
    & \mathbf{1}^T\bb = 1,~~\bb \geq 0 \\
    & (\ref{eq:milp-1}), (\ref{eq:milp-2})
\end{align}
\label{eq:opt-main-noassume}
\end{subequations}

\section{Results and Discussion}\label{sec:results}
We use the database mapping and the \gls{hag} generation frameworks (discussed in Section~\ref{sec:prelim}) to obtain the attack sequences for a component used in the \gls{cpes}. 
In this section, we discuss the results obtained for the components - (i) substation automation controller, and (ii) smart inverter.
For each component, we identify the set of \gls{attack} adversary techniques which can be executed on it.
Thereafter, we use the \gls{hag} generation framework to generate $100$ sample \gls{hag}s.
These steps are accomplished using the frameworks described in Section~\ref{sec:prelim}.
From these \gls{hag}s, we identify possible attack sequences which can be executed on the components.
In our case, we identify $397$ sequences for substation automation controller and $364$ sequences for smart inverter.
We select only those attack sequences which contains adversarial techniques included under the ``Impact'' tactic of the \gls{attack} framework.
Therefore, we shortlist the attack sequences which are meaningful in the context of creating an impact in the \gls{cpes}.

\begin{figure}[tbhp]
    \centering
    \includegraphics[width=0.47\textwidth]{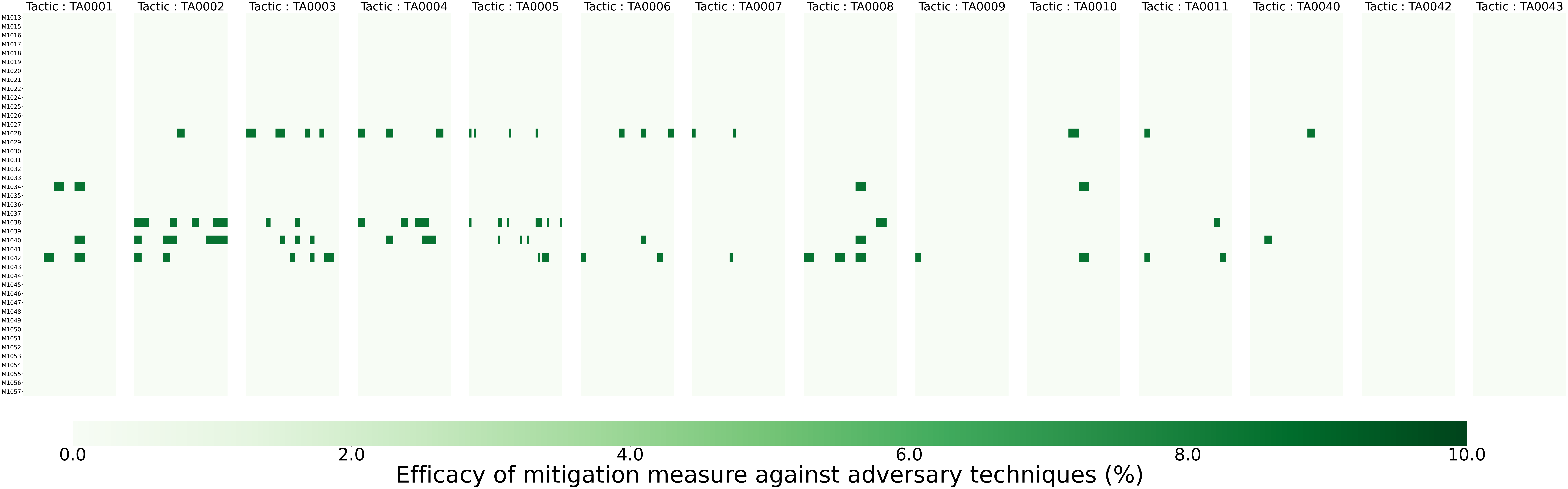}
    \includegraphics[width=0.47\textwidth]{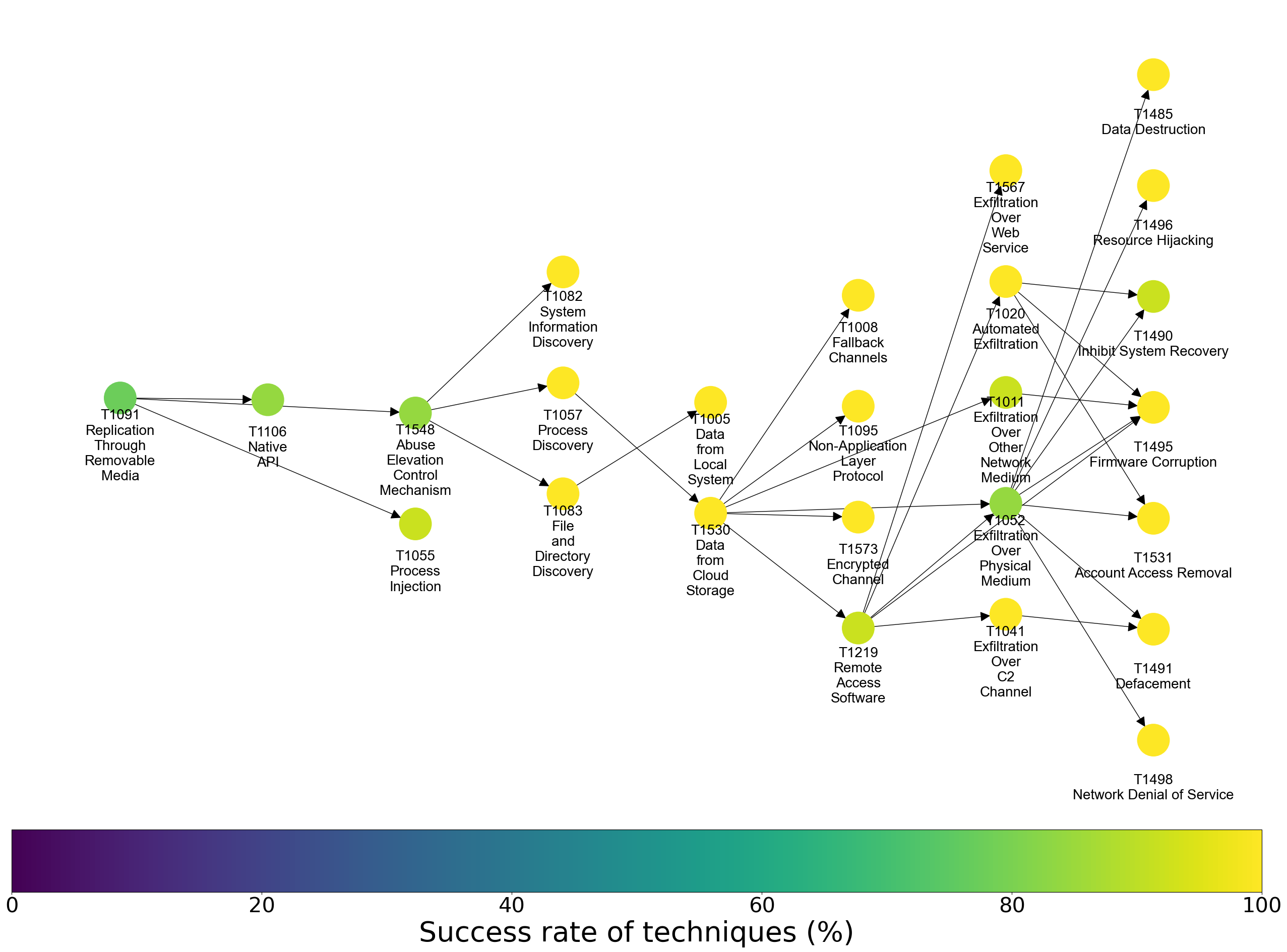}
    \caption{Matrix showing the optimal mitigation strategy for the \gls{hag} for `smart inverter' (top) and heat map of technique success rates (bottom) after implementation of the strategy.}
    \label{fig:test-hag}
\end{figure}
\begin{figure*}[tbhp]
    \centering
    \includegraphics[width=0.48\textwidth]{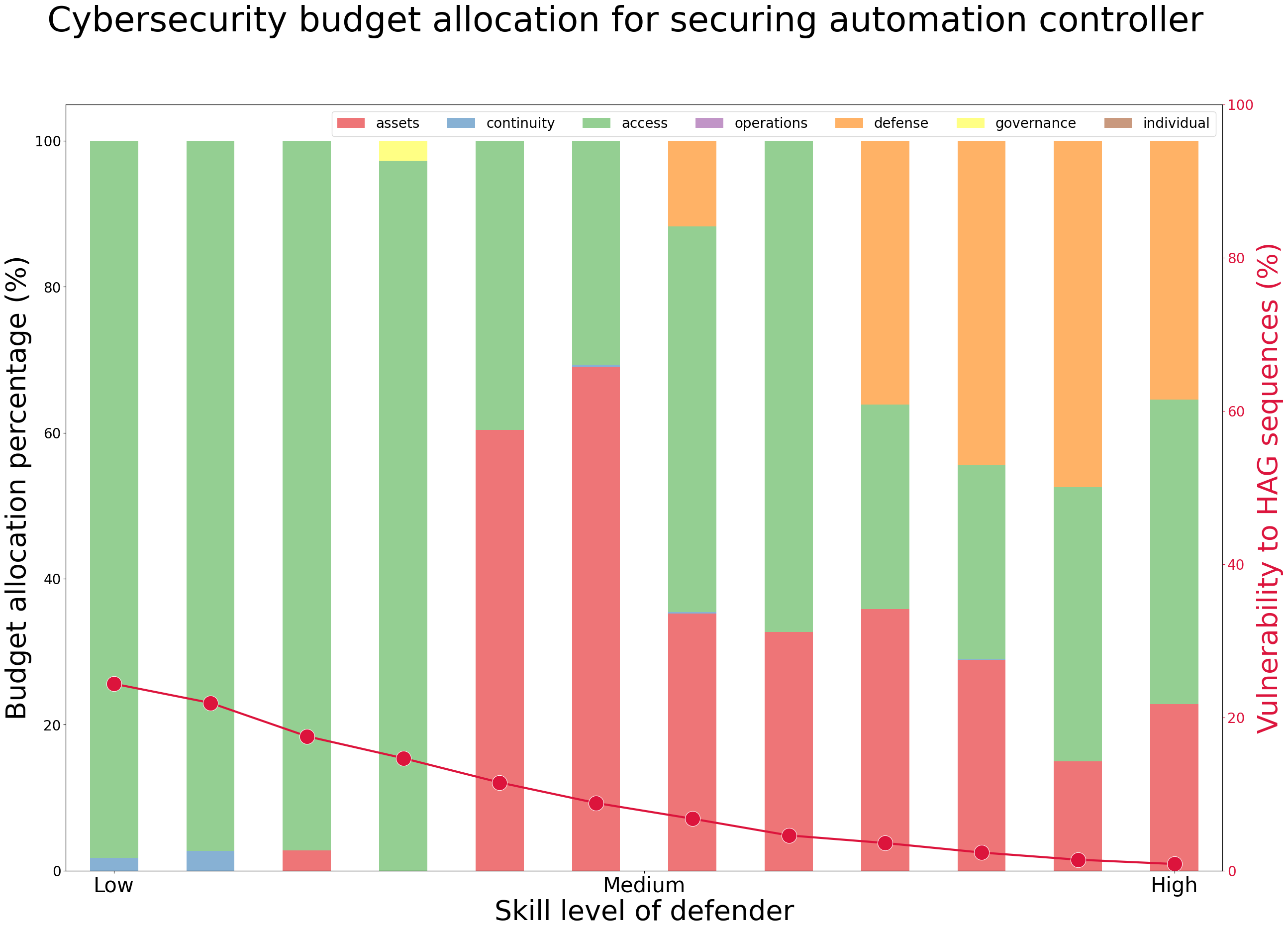}
    \includegraphics[width=0.48\textwidth]{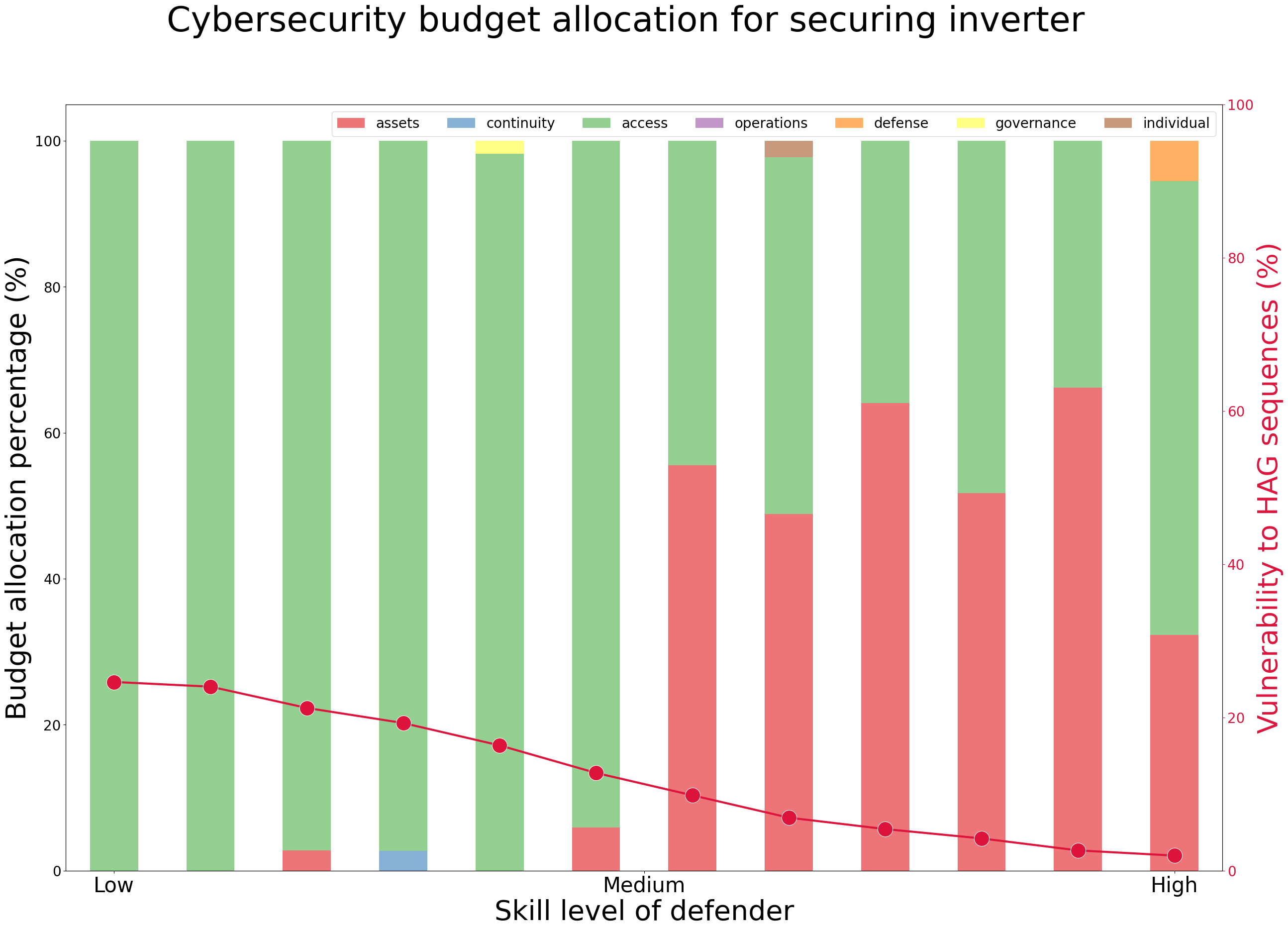}
    \caption{Optimal budget allocation in different sectors for improving cybersecurity of ``substation automation controller'' (left) and ``smart inverter'' (right). The stacked bar plots denote the percentage budget allocated to each sector and the red line plot denotes the vulnerability of the component with the prescribed budget allocation. Increased skill level of defender allows allocated budget to be efficiently used in improving the efficacy of mitigation and hence we observe reduction in vulnerability.}
    \label{fig:allocate}
\end{figure*}
\update{We assume a base case with the \gls{hag} presented in Fig.~\ref{fig:hag-inverter}, where no mitigation measures are implemented.
Therefore, all adversary techniques in the \gls{hag} have a success rate of $100\%$. 
Fig.~\ref{fig:test-hag} shows the optimal mitigation profile (top plot) and success rates of the techniques after implementing the optimal mitigation measures through a heat map on the \gls{hag} (bottom plot).
It is evident from the mitigation profile that only particular set of mitigation measures are selected.
The heat map shows impact of implementing the optimal mitigation strategy in reducing the success rate of adversary techniques in the \gls{hag}. 
We note that the optimal strategy identifies the mitigation measures which reduces the success rate of techniques such that maximum number of attack sequences are affected.
This observation can be validated from the fact that the techniques with the highest out-degree are the most influential nodes in the \gls{hag}.
These are the techniques with the lowest success rates after the mitigation measures are implemented.}

Next, we use the proposed optimization framework to partition allocated budget into various organizational sectors. 
We choose $7$ sectors and identify set of mitigation measures in each of the sectors as described in~\cite{sensors2021} -- (i) \textit{assets} includes hardware and software asset management, network infrastructure management, improving data security and privacy, (ii) \textit{continuity} sector consists of preventive strategies to continue business operations in the event of a data breach, (iii) \textit{access \& trust} deals with policies and practices for account and access management, (iv) \textit{operations} sector involves performing system risk assessment through \emph{Threat Intelligence} programs, (v) \textit{defense} sector includes mitigation measures associated with firewall implementation, (vi) \textit{governance} sector covers tasks related to audit log management and (vii) \textit{individual} category involves practices making employees aware about cybersecurity risks through training programs and performing frequent security skill evaluation.

Fig.~\ref{fig:allocate} shows the results of optimal budget allocation for two components - (i) substation automation controller and (ii) smart inverter.
We perform multiple experiments for different skill level of the defender - thereby solving an optimization problem for each skill level.
Recall that parameter $\lambda$ denotes the skill level. 
Each bar shows the partitions of the budget for a particular defender skill level. Note that the partitions sum up to $100\%$.
Further, we denote the vulnerability of the component to the \gls{hag} sequences under the optimal mitigation policy with the red dot on each bar.
This is computed using (\ref{eq:vul}) after computing the optimal value of the objective function as follows.
\begin{equation}
    \vul\left(\rD,\cM_{\sys},\delta\right) = \frac{N_{\rD}\left(\cM_{\sys}, \delta \right)}{N_{\rD}} = \frac{\mathbf{1}^T\by^\star}{N_D}
\end{equation}
where $\by^\star$ denotes the optimal value of $\by$ in (\ref{eq:opt-main-noassume}).

We note that with increases in defender skill level, the vulnerability of the component to \gls{hag} sequences reduces.
However, we notice that the budget allocation for each sector do not follow any particular trend.
This is because sectors overlap in their coverage of mitigation measures.
In the case of ``substation automation controller'', we observe that for an unskilled defender, the proposed optimization framework recommends the budget be allocated mostly towards the ``access'' sector which comprises of mitigation measures related to access management, account management and password robustness.
With a skilled defender, we observe that budget allocation gets divided to other sectors such as ``assets'' and ``defense''.
We note similar observation for ``smart inverter''---however, the budget gets divided to the ``assets'' and ``access'' sectors for more skilled defenders.

\section{Conclusion}
We propose a generalized framework which performs an optimal partitioning of a limited cybersecurity budget into various organizational sectors in order to improve the cybersecurity of a smart device or component in the \gls{cpes}.
The framework identifies the adversarial threats and possible attack sequences which can be performed to exploit cyber vulnerabilities of the component.
Thereafter, we formulate an \gls{milp} optimization problem which aims to evaluate the optimal budget partitions in order to minimize the number of highly likely attack sequences.
Though we provide results for using the framework in \gls{cpes}, the proposed methodology can be extended for any cyber-physical system.
Such a framework equips managers in an organization to formulate cybersecurity policies, allocate staff budgets in order to improve the overall security and reduce risk of \gls{apt}.

\update{In practice, a significant portion of cybersecurity budget allocation is aimed at improving software and hardware tools to prevent \gls{apt} along with hiring skilled cybersecurity personnel.
In our paper, we combine aspects of cybersecurity tools and personnel skill through the parameters of efficacy and defender skill in our simplified analytic expressions.
We plan to identify dedicated parameters which quantify these aspects in order to infuse realism in our model as part of our future work.}



\bibliography{aaai24}

\end{document}